\documentclass[a4paper,11pt]{article}
\usepackage{pos}
\usepackage{picture}
\usepackage{cancel}
\usepackage{tabularx}
\usepackage{tensor}
\usepackage{float}
\usepackage{fancyhdr}
\usepackage{xcolor}
\usepackage{scalerel}
\usepackage{relsize}
\usepackage[absolute,overlay]{textpos}
\usepackage[customcolors]{hf-tikz}
\usepackage{cals}
\usepackage{enumitem}
\setlist{nolistsep}
\usepackage{pifont}
\usepackage[export]{adjustbox} % adds left,center,right as options for \includegraphics
\usepackage[most]{tcolorbox}
\usepackage{varwidth}
\usepackage{ragged2e}
\usepackage{caption}

\graphicspath{
{./}
  {../img/}
}

\title{The confined-deconfined surface tension in SU(N) gauge theories at large N}
\author*[a,b]{Ahmed Salami}
\author[c]{Tobias Rindlisbacher}
\author[a,b]{Kari Rummukainen}

\affiliation[a]{Department of Physics,
P.O. Box 64, FI-00014 University of Helsinki, Finland}

\affiliation[b]{Helsinki Institute of Physics,
P.O. Box 64, FI-00014 University of Helsinki, Finland}

\affiliation[c]{AEC, Institute for Theoretical Physics, University of Bern, Sidlerstrasse 5, CH-3012 Bern, Switzerland}

\emailAdd{ahmed.salami@helsinki.fi}
\emailAdd{trindlis@itp.unibe.ch}
\emailAdd{kari.rummukainen@helsinki.fi}

\abstract{We present results from an investigation of the $N$-dependency of the confined-deconfined interface tension and latent heat in pure SU($N$) gauge theory at large $N$. Following Moore and Turok \cite{Moore:1996bn}, we determine the interface tension by measuring the transverse fluctuations of the phase interface on large lattices with coexisting confined and deconfined phases.  We observe unambiguously that both the interface tension and latent heat scale as $N^2$ at large $N$.
\vskip .1cm
{\footnotesize  \it Preprint:  HIP-2025-4/TH}}

\FullConference{The 41st International Symposium on Lattice Field Theory (LATTICE2024)\\
 28 July - 3 August 2024\\
Liverpool, UK\\}

\begin{document}
\maketitle
\section{Introduction}
SU($N$) gauge field theories at large $N$ provide an interesting laboratory for studying thermodynamics in strongly coupled theories, including the approach to the 't Hooft limit \cite{tHooft:1973alw} and testing duality relations \cite{Witten:1998zw,Ares:2021ntv}. In this work we study the latent heat and the interface tension of the first order confinement-deconfinement phase transition at large $N$. While the critical temperature $T_c$ (in units of the square root of the string tension) and the latent heat \cite{Lucini:2012wq,Lucini:2005vg} have been studied extensively, the interface tension remains poorly determined (see \cite{Lucini:2005vg}).

We use a method introduced by Moore and Turok already almost 30 years ago \cite{Moore:1996bn}, but which has seen limited use in the lattice community. It is based on constraining the simulation so that the configurations remain in mixed confined-deconfined state and extracting the interface tension from the surface wave spectrum of the confined-deconfined interfaces. This enables us to obtain precise critical coupling, latent heat and interface tension. Because it does not suffer from supercritical slowing down, it allows the use of large lattice volumes.  We perform simulations at $N=4$, $5$, $8$ and $10$, with inverse lattice spacings $1/(aT_c) = N_t = 5 \ldots 8$, enabling the taking of the continuum limit. In addition, we have a single lattice spacing $N_t=6$ at $N=16$.

We use the standard Wilson plaquette action~\cite{Wilson:1974sk} with periodic boundary conditions and lattice coupling $\beta = 2N/g^2$. The update algorithm is a combination of heat bath~\cite{Cabibbo:1982zn,Kennedy:1985nu} and overrelaxation~\cite{Brown:1987rra,deForcrand:2005xr} updates. The simulation program has been written using the HILA lattice field theory programming framework~\cite{HILA}, allowing for easy generation of optimized executables for various computing platforms.

\section{Lattice setup and determination of $\beta_c$}
\begin{figure}[hbt]
    \centering
    \begin{minipage}[c]{0.32\textwidth}
    \includegraphics[width=\textwidth]{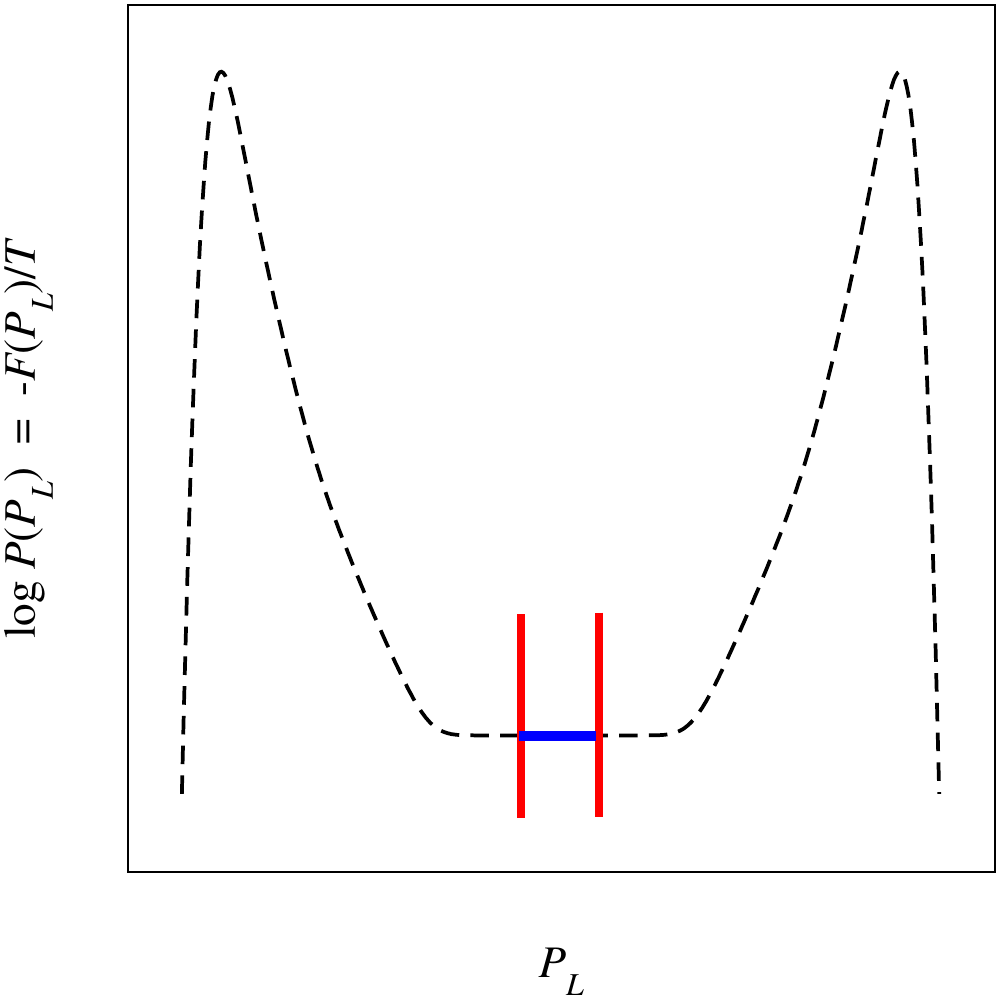}
    \end{minipage} 
    ~~~~~~~~~~
    \begin{minipage}[c]{0.45\textwidth}
    \includegraphics[width=\textwidth]{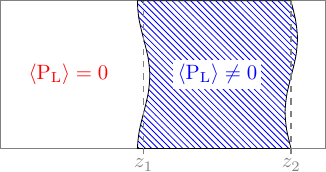}
    \end{minipage}
\caption{\emph{Left:} schematic Polyakov line $P_L$ distribution at $\beta_c$ (dashed curve), with restriction to the central region (red lines).  \emph{Right:} two-phase configuration with two planar interfaces corresponding to the central region of $P_L$ distribution.}
\label{fig:schematic}
\end{figure}

On the left panel of Fig.~\ref{fig:schematic} we show a schematic order parameter probability distribution at the critical temperature of a first order phase transition in a finite periodic volume. The bulk phase peaks have equal probability\footnote{This discussion omits the $N$-fold multiplicity of deconfined phases in SU($N$) theory. More precisely, all $N$ deconfined phases and the confined phase have equal probability at the critical point.}, and are connected by mixed phases with much reduced probability. In a large enough volume the central part becomes flat, corresponding to mixed phase with two phase interfaces as shown on the right panel. The surfaces can move with respect to each other without any cost in free energy, giving rise to the flat part of the probability distribution. 

We note that the slightly skewed shape of the probability peaks on the left panel of Fig.~\ref{fig:schematic} is due to bubble-like mixed phase configurations, see ref.~\cite{Rummukainen:2025pjj} for an application in SU(8) gauge theory.

The suppression of the mixed states is due to the interface tension $\sigma$: $P_{\textrm{min}}/P_{\textrm{max}} \approx \exp(-2A\sigma/T_c)$, where $A$ is the area of the interfaces. This property is often used to measure the interface tension from lattice simulations. Due to the usually very strongly suppressed probability of the mixed phase, various modified sampling methods can be used: multicanonical \cite{Berg:1992qua}, Wang-Landau \cite{Wang:2000fzi}, density of states \cite{Bennett:2024bhy} or Jarzynski's theorem \cite{Caselle:2016wsw}. Multicanonical methods have been used very successfully to study electroweak-like theories on the lattice, see for example \cite{Kajantie:1995kf}.
Nevertheless, these methods suffer from supercritical slowing down as the volume becomes large.

In the Moore-Turok method~\cite{Moore:1996bn,Moore:2000jw} we do not attempt to sample the full distribution.  Instead the order parameter, in our case the real part of the average Polyakov line, is restricted to a narrow range in the center of the distribution, see Fig.~\ref{fig:schematic}. This is achieved by simply rejecting updates which try to cross the barriers. The outcome is that the system remains in the mixed phase, as shown on the right panel. With a somewhat elongated lattice to $z$-direction the surfaces align along $(x,y$)-plane so that the area is minimized and the separation between them is larger.  The initial configuration is prepared so that the deconfined phase Polyakov line expectation value is on the positive real axis; due to the very large metastability the expectation value remains real over the whole simulation, justifying the use of the real part of the Polyakov line.

\begin{figure}[tb]
    \begin{minipage}[t]{0.31\linewidth}
    \vspace{0pt}
    \centering
    \includegraphics[height=1.1\linewidth,keepaspectratio,right]{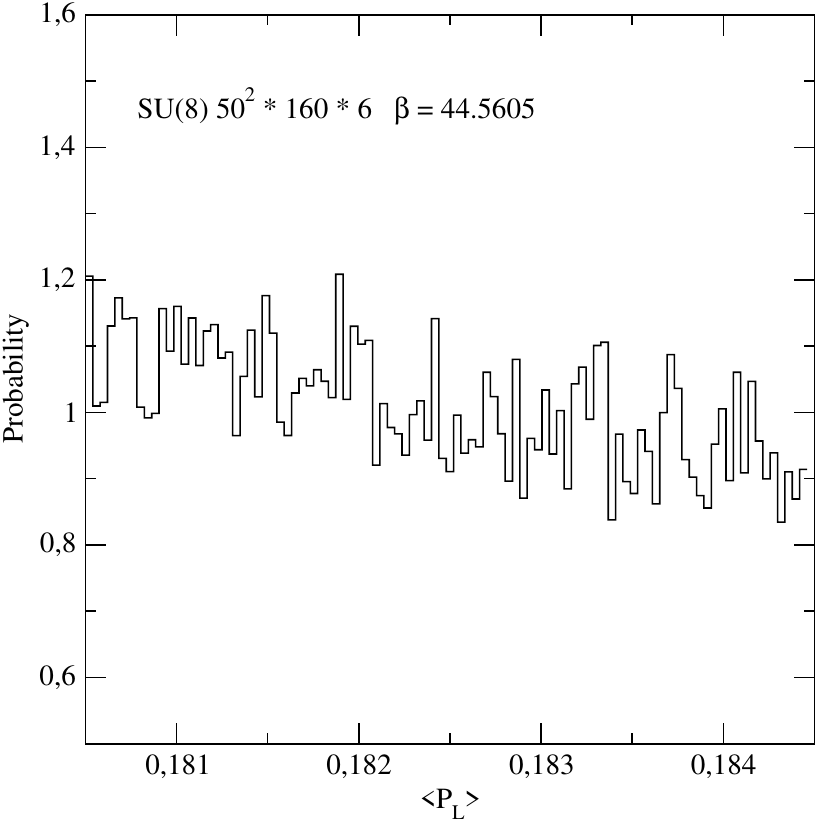}
    \end{minipage}\hfill
    \begin{minipage}[t]{0.31\linewidth}
    \vspace{0pt}
    \centering
    \includegraphics[height=1.1\linewidth,keepaspectratio,right]{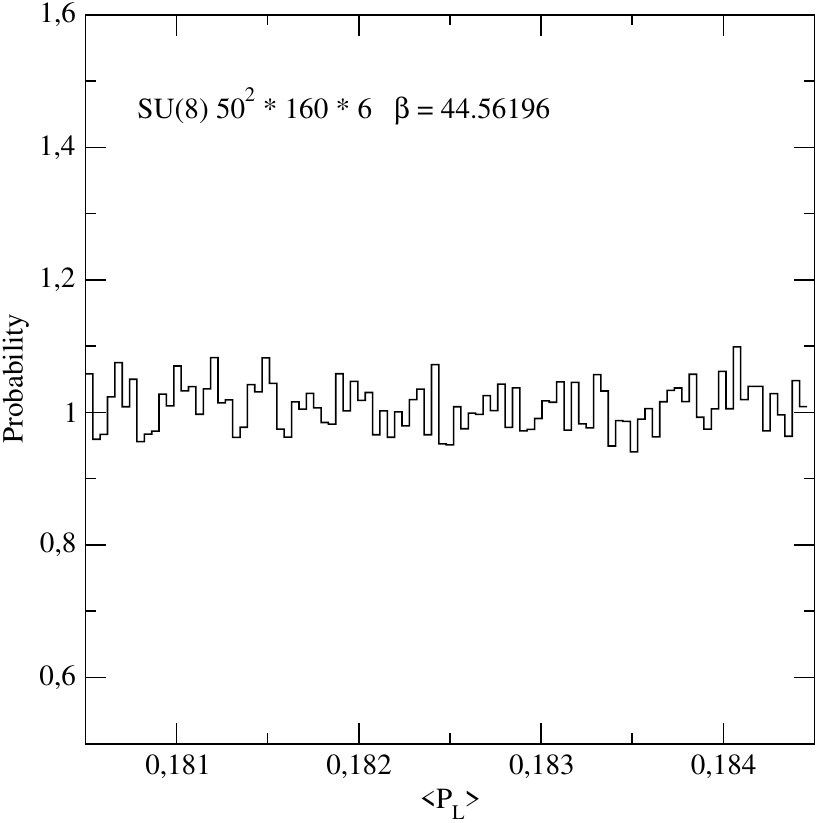}
    \end{minipage}\hfill
    \begin{minipage}[t]{0.31\linewidth}
    \vspace{0pt}
    \centering
    \includegraphics[height=1.1\linewidth,keepaspectratio,right]{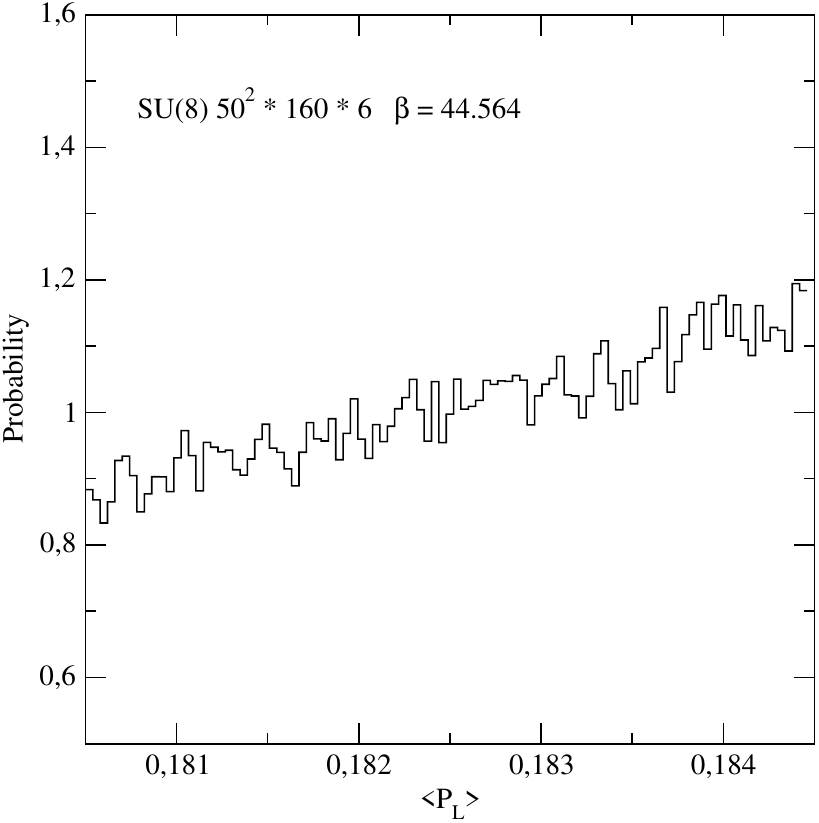}
    \end{minipage}\\[5pt]
    \caption{Restricted average Polyakov line $P_L$ 
    probability distributions of SU(8) gauge theory on a $6\times 50^2 \times 160$ lattice slightly below, at, and above the critical $\beta$.  In the simulation $P_L$ was restricted to range $0.1805 \le P_L \le 0.1845$.}
    \label{fig:histsforbetac}
\end{figure}

The critical lattice coupling $\beta_c$ can now be determined by demanding that the Polyakov line probability distribution over the restricted range becomes horizontal. An example of this is shown in Fig.~\ref{fig:histsforbetac}. 
Even a small deviation in $\beta$ away from its critical value is enough to tilt the distribution. The final value of $\beta_c$ is obtained by reweighting, and error analysis is done with the jackknife method. 

We note that this method has exponentially suppressed finite size effects. Indeed, we did not observe any finite volume systematics at large enough volumes, and we perform the measurements on the largest volumes used at each $N$ and $N_t$. The volumes all satisfy the limits $N_t \le 10 N_{x,y} \ll N_z$, with a single exception at SU(10). The simulation volumes and $\beta_c$ values are reported in Table~\ref{tab:results}.

\section{Interface tension}

Because lattices are elongated in the $z$-direction, the two interfaces between confined and deconfined phase will always be aligned along the $(x,y)$-plane. Let us describe surface of size $L^2$ as an infinitely thin sheet with height $z(x,y)$ and interface tension $\sigma$, which restricts the thermal fluctuations of $z(x,y)$. Assuming small amplitude fluctuations, it is straightforward to show \cite{Moore:1996bn} that the 2-dimensional Fourier transform of $z(x,y)$ obeys 
\begin{equation}
\langle |\hat z(n_x,n_y)|^2 \rangle = \frac{T}{\sigma L^2 (k_x^2 + k_y^2)} = \frac{T}{4\pi^2 \sigma(n_x^2+n_y^2) } ~,
\label{eq:zhat}
\end{equation}
where $k = 2\pi n/L$ is a 2-dimensional wave vector along the $(x,y)$-plane, and the expectation value is over the ensemble of configurations.
In realistic field theory the interface is not infinitely thin; however, in the limit of very long wavelength fluctuations ($k_i\rightarrow 0$) the assumption of small amplitude fluctuations compared to wavelength becomes valid and Eq.~(\ref{eq:zhat}) can be used.

In our case the location of the interface $z(x,y)$ can be obtained from the Polyakov line field, $P_L(x,y,z)$:
for each configuration and for each $(x,y)$ we search for two values (for two interfaces) of $z$ where $P_L(x,y,z)$ crosses a threshold value of one half of the deconfined phase expectation value.

\begin{figure}[htb]
\begin{minipage}[t]{0.47\linewidth}
\vspace{0pt}
\centering
\includegraphics[height=1.1\linewidth,keepaspectratio,right]{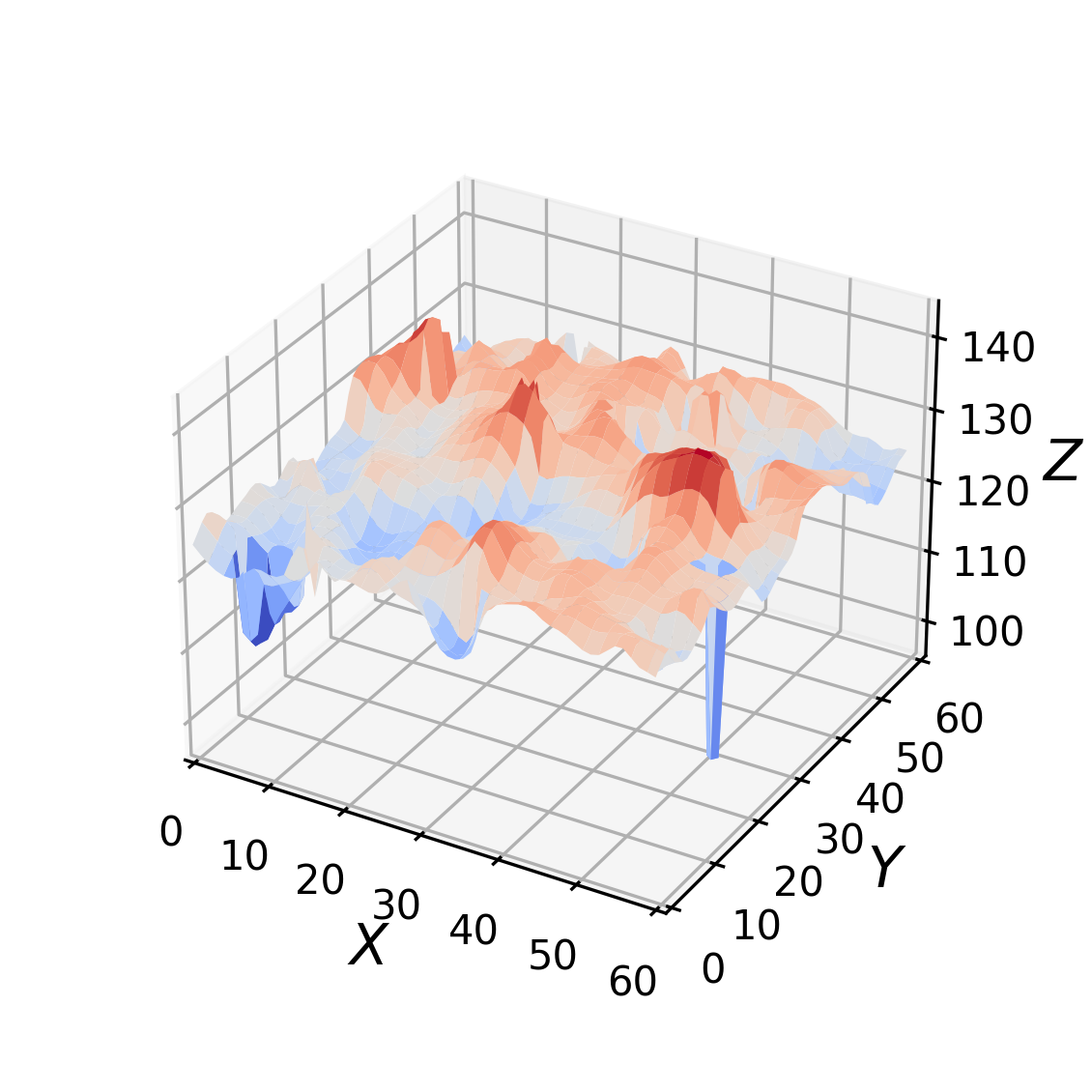}
\end{minipage}\hfill
\begin{minipage}[t]{0.47\linewidth}
\vspace{0pt}
\centering
\includegraphics[height=1.1\linewidth,keepaspectratio,right]{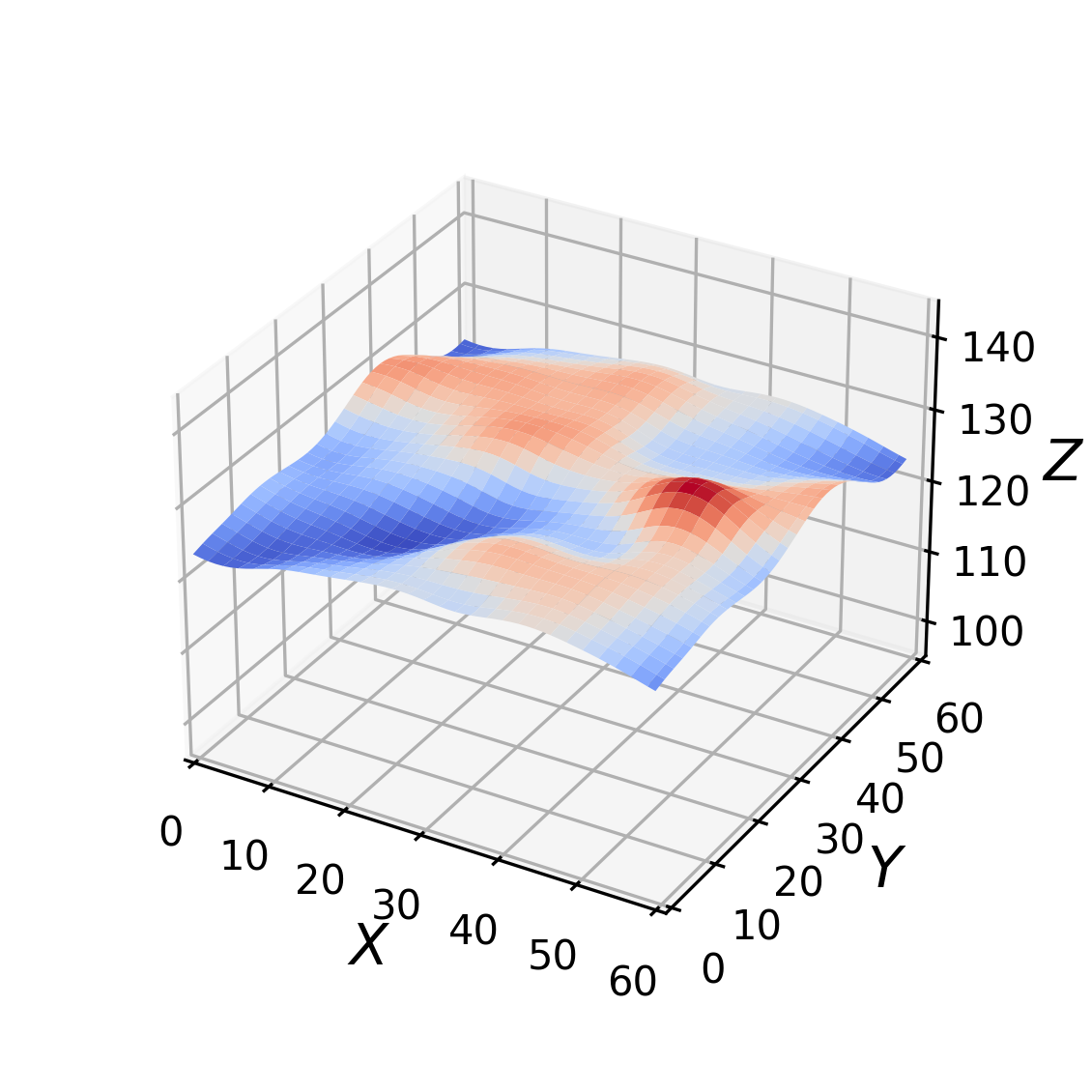}
\end{minipage}\\[5pt]
\caption{Capturing the confined-deconfined interface from a coexisting-phase-restricted simulation of SU(10) theory on a $60^2\times 240\times 6$ lattice at $\beta=69.9225$, after 10 (left) and 70 (right) smearing steps. Dislocations still present at 10 smearing steps disappear at sufficiently large number of steps.}
\label{fig:smearing_comp}
\end{figure}

However, bare $P_L$ is a very strongly fluctuating quantity and it is not usable by itself to pin down the interface.  It can be regulated by recursively smearing $P_L \rightarrow P_L^{(n_\text{smear})}$ with smearing kernel $S$:
\begin{equation}
P_L^{(n+1)}(\bar x) = \sum_{\bar{y}} S(\bar x,\bar y) P_L^{(n)}(\bar y),  ~~~~~~
S(\bar x,\bar y) = \frac{1}{1+6\,\rho}\left(\delta_{\bar x,\bar y}+\rho\sum_{i=1}^{3}(\delta_{\bar x+\hat{i},\bar y}+\delta_{\bar x -\hat{i},\bar y})\right)\ .\label{eq:smearingkernel}
\end{equation}
The smearing acts as an UV filter. Too few smearing steps do not reduce the bulk fluctuations sufficiently, leading to spurious dislocations in interface locations, as shown on the left panel in Fig.~\ref{fig:smearing_comp}. Increasing the smearing level makes the dislocations disappear while preserving the long-wavelength structure, unless the number of smearing steps grows excessive.

\begin{figure}[tb]
    \centering
    \hfill
    \begin{minipage}[t]{0.47\linewidth}
    \includegraphics[width=\textwidth]{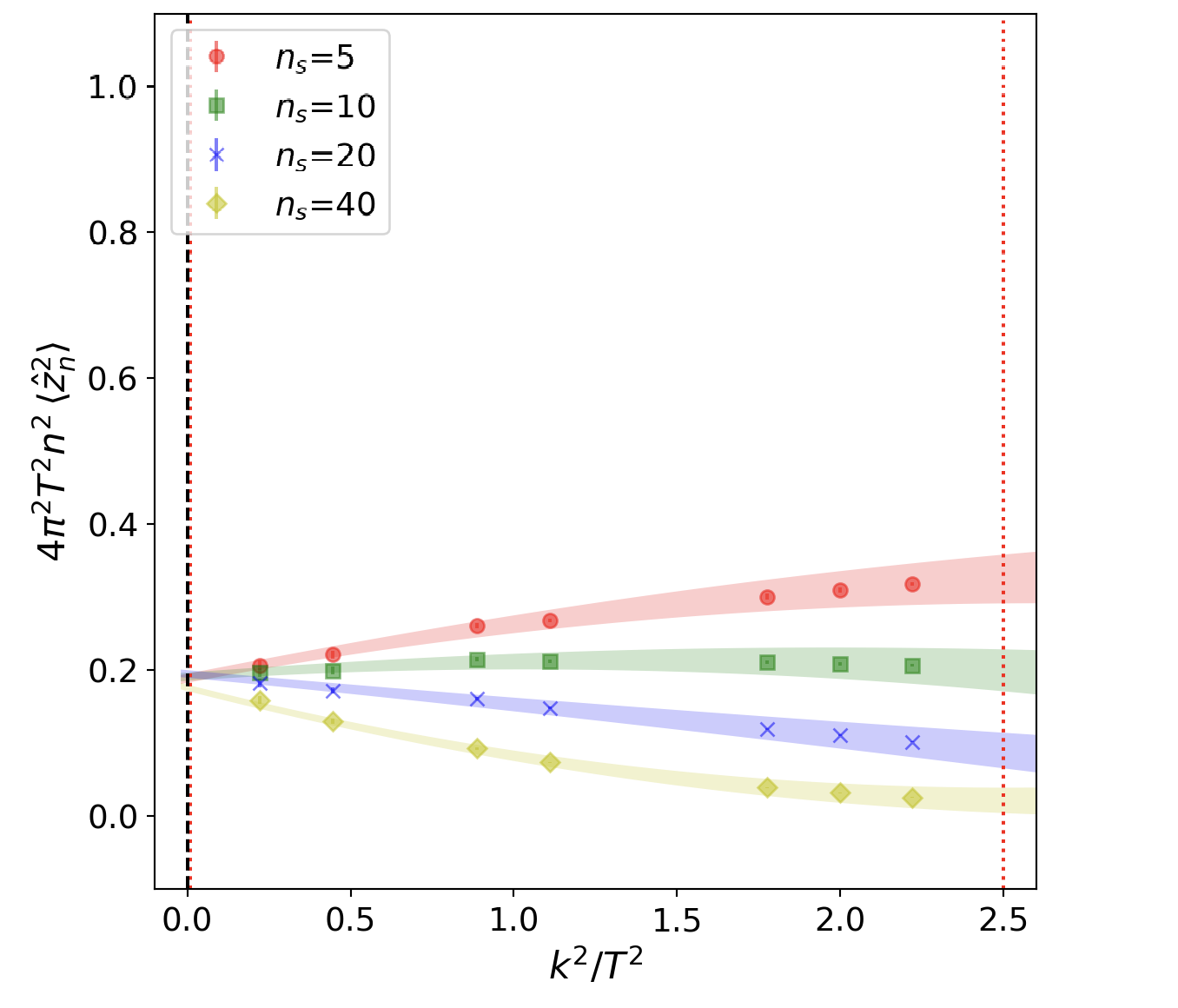}
    \end{minipage}\hfill
    \begin{minipage}[t]{0.47\linewidth}
    \includegraphics[width=\textwidth]{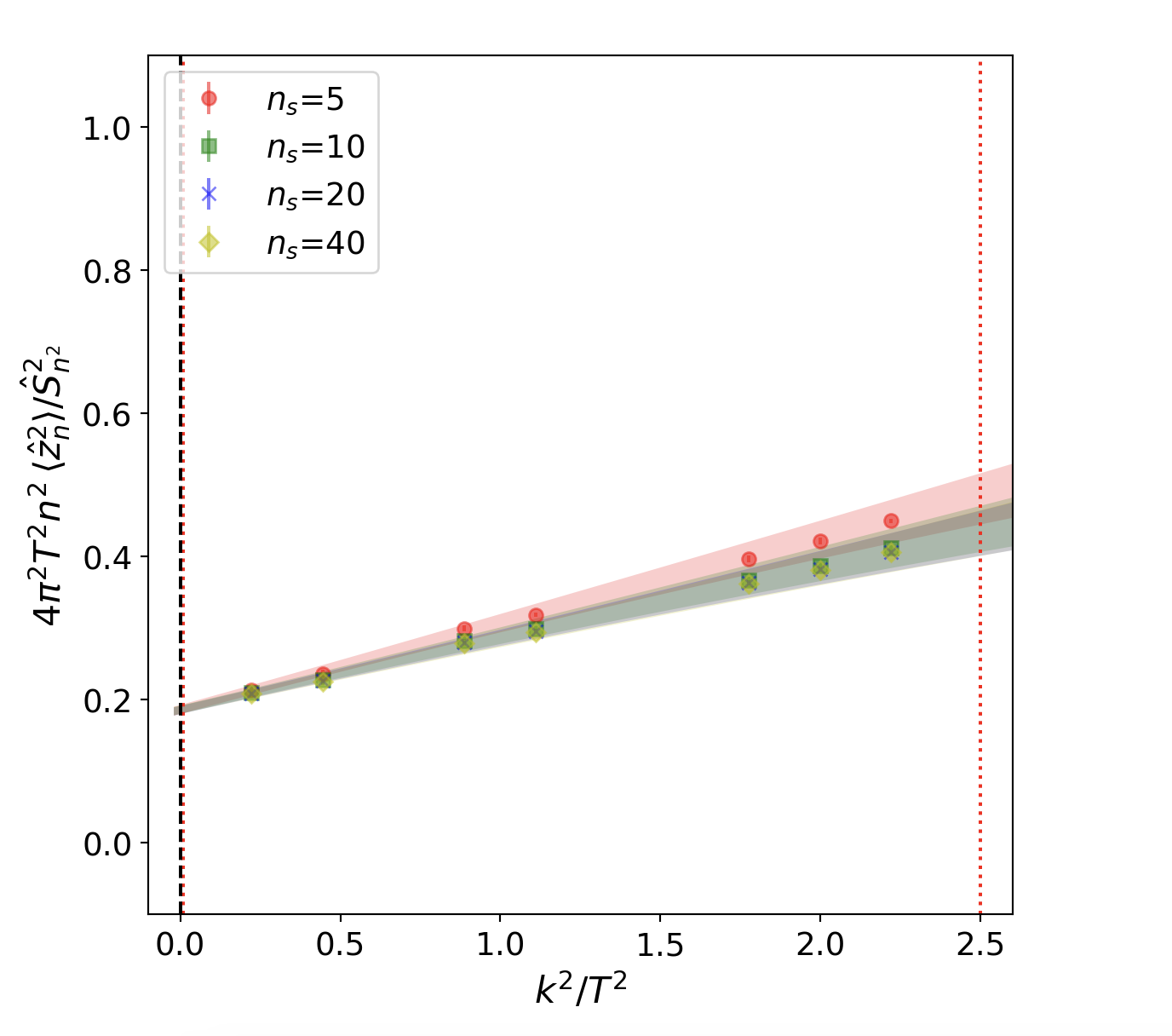}
    \end{minipage}\hfill\hfill
    \caption{\emph{Left:} extrapolation of the interface Fourier modes to $k\rightarrow 0$ limit at different number of smearing steps for SU(16) at $\beta=179.855$ on a $40^2\times 160 \times 6$ lattice.
    \emph{Right:} same after kernel correction. Data collapses on a single curve, except for the lowest smearing level. The $k\rightarrow 0$ extrapolation gives the value of $1/\sigma$.}
    \label{fig:fouriermodes}
\end{figure}

An example of the behaviour of the lowest $k$ modes at different smearing levels is shown on the left panel in Fig.~\ref{fig:fouriermodes}. Clearly, the smearing decreases the magnitude of the higher $k$ Fourier modes, but nevertheless the extrapolation to $k\rightarrow 0 $ remains stable with a large range of smearing levels.  The result of the extrapolation gives the inverse of $\sigma$.
From Fig.~\ref{fig:fouriermodes} we also see that at small smearing levels the magnitude of $n^2 |\hat z_n|^2$
grows as $n^2$ increases, i.e. there are more fluctuations at short wavelengths than predicted by interface fluctuations.  This is due to contamination by bulk fluctuations of the order parameter.

The control over the smearing is illustrated by the fact that the effect of the smearing on the expectation values of Fourier modes can be calculated semi-analytically. To see this, let 
\begin{equation}
A(z_0(x,y),z_1(x,y),z)=
\begin{cases}
A_0 \quad\text{if} & z_0(x,y) \leq z < z_1(x,y)\\
0 \quad\text{else} & \\
\end{cases}\ ,
\end{equation}
describe the shape of the volume occupied by the deconfined phase, neglecting bulk fluctuations, so that $A_0=\left|\langle P_L \rangle_{\text{deconf.}}\right|$ is constant. We now Fourier transform in $z$:
\begin{equation}
\hat{A}(z_0(x,y),z_1(x,y),k_z)=\frac{i\,A_0}{\sqrt{2\,\pi}\,k_z}\,\left(e^{i\,k_z z_0(x,y)}-e^{i \,k_z z_1(x,y)}\right)\ .
\end{equation}
Since in our setup, the two phases occupy about the same space in z-direction and $A(z)\in\left[0,A_0\right]$, the $k_z=0$ mode should dominate and we can expand:
\begin{equation}
\hat{A}(z_0(x,y),z_1(x,y),k_z)=\frac{A_0}{\sqrt{2\pi}}\left(z_1(x,y)-z_0(x,y)\right) + \mathcal{O}(k_z)\ .
\end{equation}
To leading order in $k_z$, the Fourier transform of $\tilde{A}(z_0(x,y),z_1(x,y),k_z)$ in x and y, is therefore given by
\begin{equation}
\hat{A}(k_x,k_y,k_z)=\frac{A_0}{\sqrt{2\pi}}\left(\hat{z}_1(k_x,k_y)-\hat{z}_0(k_x,k_y)\right) + \mathcal{O}(k_z)\ .
\end{equation}
After $n_s$ smearing steps, the spectrum of the 3D function $A(x,y,z)$ is
\begin{equation}
\hat{A}^{(n_s)}(k_x,k_y,k_z) = \hat{S}^{n_s}(k_x,k_y,k_z)\,\hat{A}(k_x,k_y,k_z)\ ,
\end{equation}
and in the limit $k_z\to 0$ one therefore has
\begin{equation}
\hat{z}^{(n_s)}(k_x,k_y)=\hat{S}^{n_s}(k_x,k_y,0)\,\hat{z}(k_x,k_y)\ .
\end{equation}
This enables us to undo the effect of the smearing on the spectrum by dividing by the square of Fourier transformed smearing kernel, as shown on the right panel of Fig.~\ref{fig:fouriermodes}. At high enough smearing levels the kernel corrected measurements collapse on a single curve. For large $N$, this happens relatively quickly; at smaller $N$ the number of required smearing steps increases and the range in momenta over which the collapse occurs decreases, requiring larger lattices. 

\begin{figure}
    \centering
    \begin{minipage}[b]{0.49\textwidth}
    \includegraphics[width=\textwidth]{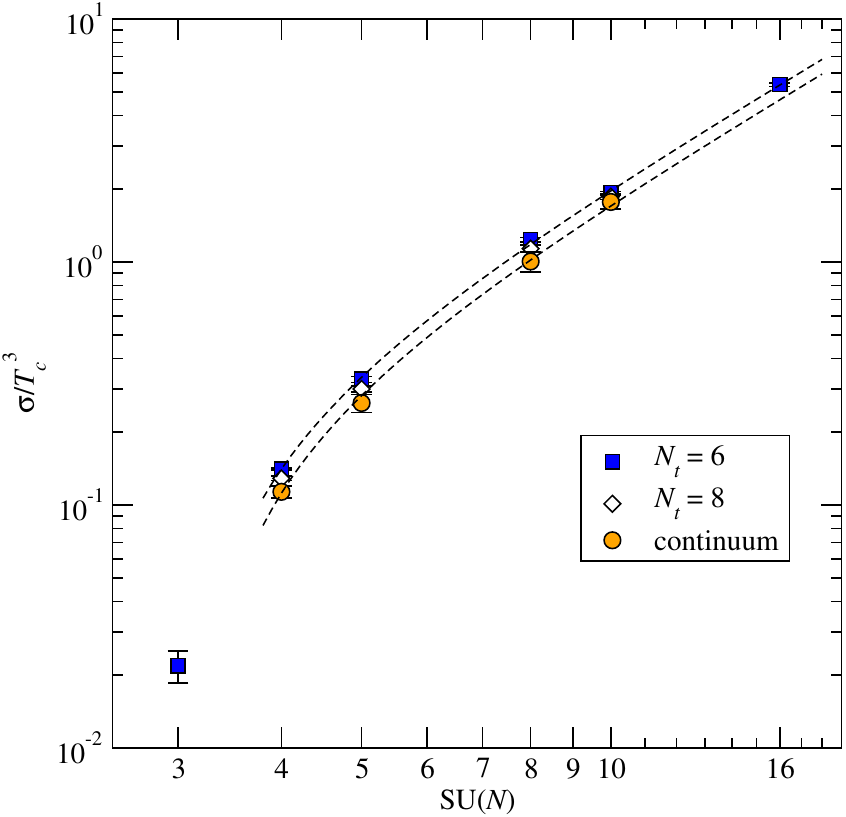}
    \end{minipage} 
    \hfill
    \begin{minipage}[b]{0.49\textwidth}
    \includegraphics[width=\textwidth]{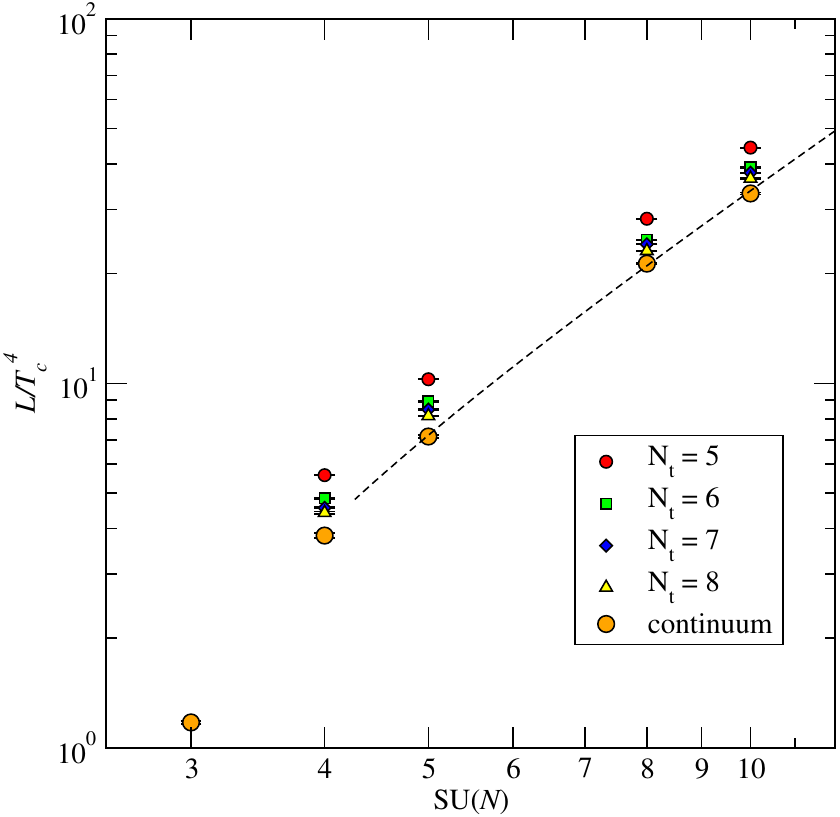}
    \end{minipage}
\caption{\emph{Left:} interface tension $\sigma/T_c^3$ as a function of $N$ shown on a log-log plot. Dashed lines are large-$N$ fits to $N_t=6$ and continuum limit results.  The SU(3) $N_t=6$ point is from ref.~\cite{Iwasaki:1993qu} and not used in the fit. \emph{Right:} latent heat $L/T_c^4$, with the continuum large-$M$ fit to $N\ge 5$ as a dashed line.  The SU(3) continuum value is from \cite{Giusti:2025fxu}.}
\label{fig:result}
\end{figure}

We performed the interface tension analysis on $N_t = 6$ and $8$ lattices, and extrapolated to continuum linearly in $a^2$.  The results are shown in Table~\ref{tab:results} and on the left panel of Fig.~\ref{fig:result}.  
At $N_t=6$ we did a measurement with SU(16), in order to approach further into the large $N$ region.  Both the $N_t=6$ and continuum limit data can be fitted with $N^2 + \text{\emph{const.}}$ expression; the continuum limit fit is
\begin{equation}
    \frac{\sigma}{T_c^3} = 0.0189(11) N^2 - 0.190(19).
\end{equation}
This result is $\sim 40\%$ larger than the non-continuum limit $N_t = 5$ result presented in ref.~\cite{Lucini:2005vg}.

\begin{table}[h]
    \small
    \centering
    \begin{tabular}{l | c | c | c | c | c | c }
        $N_c$ & $N_t$ & $N_{x,y}$ & $N_z$ & $\beta_c$ & $\sigma/T_c^3$ & $L/T_c^4$ \\
	\hline \hline
        4 & 5 & 80 & 320 & 10.637765(83) &             & 5.6027(94) \\
          & 6 & 80 & 200 & 10.79191(11)  & 0.1405(15)  & 4.836(13) \\
          & 7 & 100 & 400 & 10.94215(25) &             & 4.554(19) \\
          & 8 & 140 & 400 & 11.08443(35) & 0.1286(27)  & 4.415(35) \\
          & $\infty$& &     &            &  0.1132(66)  & 3.825(60) \\
        \hline
        5 & 5 & 80 & 240 & 16.87628(15)  &             & 10.2667(75) \\
          & 6 & 160 & 480 & 17.11085(11) & 0.3285(92)  & 8.919(33) \\
          & 7 & 100 & 400 & 17.34267(21) &             & 8.479(20) \\
          & 8 & 120 & 360 & 17.56119(12) & 0.2997(84)  & 8.143(21) \\
        & $\infty$& &       &             & 0.263(22)  & 7.148(65) \\

        \hline
        8 & 5 & 50 & 180 & 43.98229(34)  &           & 28.286(19) \\
          & 6 & 80 & 240 & 44.56196(48)  & 1.235(24) & 24.751(21) \\
          & 7 & 60 & 240 & 45.13525(76)  &           & 24.080(45) \\
          & 8 & 100 & 400 & 45.67833(49) & 1.134(39) & 23.100(47) \\
        & $\infty$& &       &            & 1.004(94) & 21.304(98) \\
        \hline
        10 & 5 & 50 & 200 & 69.03126(38) &           & 44.356(43) \\
           & 6 & 60 & 240 & 69.92349(80) & 1.926(25) & 39.152(91) \\
           & 7 & 60 & 240 & 70.80712(94) &           & 37.844(50) \\
           & 8 & 80 & 280 & 71.64759(94) & 1.856(49) & 36.500(68) \\
        & $\infty$& &     &              & 1.77(12)  & 33.21(19) \\
        \hline
        16 & 6 & 40 & 160 & 179.8508(20) & 5.364(89) & 
    \end{tabular}
    \caption{Critical values $\beta_c$ of the inverse coupling $\beta=2\,N_c/g_0^2$, interface tension $\sigma$, and latent heat $L$ for different numbers of colors $N_c$ and the lattice size used $N_t\times N^2_{x,y}\times N_z$. 
    The continuum limit extrapolations are labeled with $N_t = \infty$.}
    \label{tab:results}
\end{table}

\section{Latent heat}

Latent heat can be calculated from the discontinuity of the derivative of the free energy density
\begin{equation}
    L = \Delta \frac{\mathrm{d}}{\mathrm{d}T}\frac F V = \frac {T_c^2}{V} \Delta \frac{\mathrm{d}}{\mathrm{d}T} \ln Z 
\end{equation}
where $\Delta$ is discontinuity at $T=T_c$.  On the lattice this can be expressed in terms of the
discontinuity of the plaquette expectation value $\Delta \langle U_\Box \rangle$ at $\beta = \beta_c$:
\begin{equation}
    \frac L{T_c^4} = -N_t^4 { \frac{\mathrm{d} \beta}{\mathrm{d} \ln a} } {\Delta \langle U_\Box \rangle}.
    \label{eq:latent}
\end{equation}
The inverse lattice $\beta$-function ${\mathrm{d} \beta}/{\mathrm{d} \ln a}$ is often evaluated using string tension or gradient flow for scale setting.  However, since we have accurate measurements of $\beta_c$ as functions of $N_t = 1/(a T_c)$, we use $T_c$ as the fixed physical scale. Because the data is discrete, in order to evaluate the derivative we interpolate the measurements to continuous $N_t = 1/(aT_c)$. We use the ansatz \cite{Lucini:2005vg}
\begin{equation}
    a T_c = \frac 1 {N_t} = \frac 1 {N_{t,0}} \exp\left[ -\frac{12 \pi^2}{11 N^2} (\beta_c(N_t) - \beta_c(N_{t,0})) + \sum_{i=1}^3 c_i z^i\right],
    \label{eq:interp}
\end{equation}
where $z = N^2[1/{\beta_c(N_t)} - 1/{\beta_c(N_{t,0})}]$ and $c_i$ are fit parameters.  We choose the reference value $N_{t,0} = 6$, the final result does not depend on this choice. The leading behaviour of the function is motivated by the asymptotic 1-loop $\beta$-function. 

The measurements of $\beta_c$ have very small errors, and it is necessary to have 3 fit parameters for each SU($N$), saturating the degrees of freedom of the fit. 
Using function (\ref{eq:interp}) it is straightforward to evaluate the derivative.

We measure the plaquette discontinuity by separate simulations at $\beta_c$, prepared so that the system is fully in the confined or the deconfined phase, and using large enough volumes so that the system remains in the same phase throughout the simulation. This allows us to obtain the latent heat for each $N_t$ and $N$. The results are shown on the right panel in Fig.~\ref{fig:result}.  We extrapolate to continuum linearly in $a^2$, using $N_t = 6,7,8$ values. $N_t = 5$ deviates from the linear fit and we exclude it here. 
For $N \ge 5$, the continuum limit is well described by the large-$N$ fit 
\begin{equation}
  \frac{L}{T_c^4} = 0.354(2) N^2 - 1.65(10),  ~~~~~   N \ge 5,
\end{equation}
also shown in Fig.~\ref{fig:result}.  In ref.~\cite{Lucini:2005vg} the large $N$
continuum result was presented in form 
\begin{equation}
    \frac{L^{1/4}}{N^{1/2} T_c} = A + \frac{B}{N^2}
\end{equation}
with $A= 0.766(40)$ and $B=-0.34(1.60)$; our result expressed in this form is 
$A = 0.7731(12)$ and $B=-0.959(58)$, compatible with ref.~\cite{Lucini:2005vg} but with significantly smaller errors.

\section{Conclusion}
We have used the mixed-phase method to measure the confined-deconfined phase transition critical coupling, latent heat and interface tension in SU($N$) gauge theories with $N=4$, $5$, $8$, $10$ and $16$, obtaining continuum limit for the latent heat and the interface tension for $N \le 10$.  This gives us the large-$N$
results $L/T_c^4 = 0.354(2) N^2 - 1.65(10)$ and $\sigma/T_c^3 = 0.0189(11) N^2 - 0.190(19)$.  The method presents as a consistent and reliable means to studying stronger transitions in general.

\acknowledgments

The authors thank CSC - IT Center for Science, Finland, for computational resources.
K.R. and A.S. acknowledge support by the Research Council of Finland grant 354572 and European Research Council grant 101142449. T.R. acknowledges support from the Swiss National Science Foundation (SNSF) through the grant no.~210064.

\bibliographystyle{JHEP}
\bibliography{suN}

\providecommand{\href}[2]{#2}\begingroup\raggedright\begin{thebibliography}{10}

\bibitem{Moore:1996bn}
G.D.~Moore and N.~Turok, \emph{{Classical field dynamics of the electroweak
  phase transition}},
  \href{https://doi.org/10.1103/PhysRevD.55.6538}{\emph{Phys. Rev. D}
  {\bfseries 55} (1997) 6538}
  [\href{https://arxiv.org/abs/hep-ph/9608350}{{\ttfamily hep-ph/9608350}}].

\bibitem{tHooft:1973alw}
G.~'t~Hooft, \emph{{A Planar Diagram Theory for Strong Interactions}},
  \href{https://doi.org/10.1016/0550-3213(74)90154-0}{\emph{Nucl. Phys. B}
  {\bfseries 72} (1974) 461}.

\bibitem{Witten:1998zw}
E.~Witten, \emph{{Anti-de Sitter space, thermal phase transition, and
  confinement in gauge theories}},
  \href{https://doi.org/10.4310/ATMP.1998.v2.n3.a3}{\emph{Adv. Theor. Math.
  Phys.} {\bfseries 2} (1998) 505}
  [\href{https://arxiv.org/abs/hep-th/9803131}{{\ttfamily hep-th/9803131}}].

\bibitem{Ares:2021ntv}
F.R.~Ares, O.~Henriksson, M.~Hindmarsh, C.~Hoyos and N.~Jokela,
  \emph{{Effective actions and bubble nucleation from holography}},
  \href{https://doi.org/10.1103/PhysRevD.105.066020}{\emph{Phys. Rev. D}
  {\bfseries 105} (2022) 066020}
  [\href{https://arxiv.org/abs/2109.13784}{{\ttfamily 2109.13784}}].

\bibitem{Lucini:2012wq}
B.~Lucini, A.~Rago and E.~Rinaldi, \emph{{SU($N_c$) gauge theories at
  deconfinement}},
  \href{https://doi.org/10.1016/j.physletb.2012.04.070}{\emph{Phys. Lett. B}
  {\bfseries 712} (2012) 279}
  [\href{https://arxiv.org/abs/1202.6684}{{\ttfamily 1202.6684}}].

\bibitem{Lucini:2005vg}
B.~Lucini, M.~Teper and U.~Wenger, \emph{{Properties of the deconfining phase
  transition in SU(N) gauge theories}},
  \href{https://doi.org/10.1088/1126-6708/2005/02/033}{\emph{JHEP} {\bfseries
  02} (2005) 033} [\href{https://arxiv.org/abs/hep-lat/0502003}{{\ttfamily
  hep-lat/0502003}}].

\bibitem{Wilson:1974sk}
K.G.~Wilson, \emph{{Confinement of Quarks}},
  \href{https://doi.org/10.1103/PhysRevD.10.2445}{\emph{Phys. Rev. D}
  {\bfseries 10} (1974) 2445}.

\bibitem{Cabibbo:1982zn}
N.~Cabibbo and E.~Marinari, \emph{{A New Method for Updating SU(N) Matrices in
  Computer Simulations of Gauge Theories}},
  \href{https://doi.org/10.1016/0370-2693(82)90696-7}{\emph{Phys. Lett. B}
  {\bfseries 119} (1982) 387}.

\bibitem{Kennedy:1985nu}
A.D.~Kennedy and B.J.~Pendleton, \emph{{Improved Heat Bath Method for Monte
  Carlo Calculations in Lattice Gauge Theories}},
  \href{https://doi.org/10.1016/0370-2693(85)91632-6}{\emph{Phys. Lett. B}
  {\bfseries 156} (1985) 393}.

\bibitem{Brown:1987rra}
F.R.~Brown and T.J.~Woch, \emph{{Overrelaxed Heat Bath and Metropolis
  Algorithms for Accelerating Pure Gauge Monte Carlo Calculations}},
  \href{https://doi.org/10.1103/PhysRevLett.58.2394}{\emph{Phys. Rev. Lett.}
  {\bfseries 58} (1987) 2394}.

\bibitem{deForcrand:2005xr}
P.~de~Forcrand and O.~Jahn, \emph{{Monte Carlo overrelaxation for SU(N) gauge
  theories}},  in \emph{QCD and Numerical Analysis III}, (Berlin, Heidelberg),
  pp.~67--73, Springer Berlin Heidelberg, 3, 2005,
  \href{https://doi.org/10.1007/3-540-28504-0_6}{DOI}
  [\href{https://arxiv.org/abs/hep-lat/0503041}{{\ttfamily hep-lat/0503041}}].

\bibitem{HILA}
``Hila lattice simulation framework.'' \url{https://github.com/CFT-HY/HILA}.

\bibitem{Rummukainen:2025pjj}
K.~Rummukainen, R.~Sepp\"a and D.J.~Weir, \emph{{Resolving the critical bubble
  in SU(8) deconfinement transition}},  in \emph{{41st International Symposium
  on Lattice Field Theory}}, 1, 2025
  [\href{https://arxiv.org/abs/2501.17593}{{\ttfamily 2501.17593}}].

\bibitem{Berg:1992qua}
B.A.~Berg and T.~Neuhaus, \emph{{Multicanonical ensemble: A New approach to
  simulate first order phase transitions}},
  \href{https://doi.org/10.1103/PhysRevLett.68.9}{\emph{Phys. Rev. Lett.}
  {\bfseries 68} (1992) 9}
  [\href{https://arxiv.org/abs/hep-lat/9202004}{{\ttfamily hep-lat/9202004}}].

\bibitem{Wang:2000fzi}
F.~Wang and D.P.~Landau, \emph{{Efficient, Multiple-Range Random Walk Algorithm
  to Calculate the Density of States}},
  \href{https://doi.org/10.1103/PhysRevLett.86.2050}{\emph{Phys. Rev. Lett.}
  {\bfseries 86} (2001) 2050}
  [\href{https://arxiv.org/abs/cond-mat/0011174}{{\ttfamily
  cond-mat/0011174}}].

\bibitem{Bennett:2024bhy}
E.~Bennett, B.~Lucini, D.~Mason, M.~Piai, E.~Rinaldi and D.~Vadacchino,
  \emph{{The density of states method for symplectic gauge theories at finite
  temperature}},  \href{https://arxiv.org/abs/2409.19426}{{\ttfamily
  2409.19426}}.

\bibitem{Caselle:2016wsw}
M.~Caselle, G.~Costagliola, A.~Nada, M.~Panero and A.~Toniato,
  \emph{{Jarzynski\textquoteright{}s theorem for lattice gauge theory}},
  \href{https://doi.org/10.1103/PhysRevD.94.034503}{\emph{Phys. Rev. D}
  {\bfseries 94} (2016) 034503}
  [\href{https://arxiv.org/abs/1604.05544}{{\ttfamily 1604.05544}}].

\bibitem{Kajantie:1995kf}
K.~Kajantie, M.~Laine, K.~Rummukainen and M.E.~Shaposhnikov, \emph{{The
  Electroweak phase transition: A Nonperturbative analysis}},
  \href{https://doi.org/10.1016/0550-3213(96)00052-1}{\emph{Nucl. Phys. B}
  {\bfseries 466} (1996) 189}
  [\href{https://arxiv.org/abs/hep-lat/9510020}{{\ttfamily hep-lat/9510020}}].

\bibitem{Moore:2000jw}
G.D.~Moore and K.~Rummukainen, \emph{{Electroweak bubble nucleation,
  nonperturbatively}},
  \href{https://doi.org/10.1103/PhysRevD.63.045002}{\emph{Phys. Rev. D}
  {\bfseries 63} (2001) 045002}
  [\href{https://arxiv.org/abs/hep-ph/0009132}{{\ttfamily hep-ph/0009132}}].

\bibitem{Iwasaki:1993qu}
Y.~Iwasaki, K.~Kanaya, L.~Karkkainen, K.~Rummukainen and T.~Yoshie,
  \emph{{Interface tension in quenched QCD}},
  \href{https://doi.org/10.1103/PhysRevD.49.3540}{\emph{Phys. Rev. D}
  {\bfseries 49} (1994) 3540}
  [\href{https://arxiv.org/abs/hep-lat/9309003}{{\ttfamily hep-lat/9309003}}].

\bibitem{Giusti:2025fxu}
L.~Giusti, M.~Hirasawa, M.~Pepe and L.~Virz\`\i{}, \emph{{A precise study of
  the SU(3) Yang-Mills theory across the deconfinement transition}},
  \href{https://arxiv.org/abs/2501.10284}{{\ttfamily 2501.10284}}.

\end{thebibliography}\endgroup

\end{document}